\documentstyle[amssymb,preprint, aps]{revtex}

\begin{document}
\draft
\title{Electron Spin Relaxation in a Semiconductor Quantum Well }
\author{Vadim I. Puller, Lev G. Mourokh and Norman J.M. Horing}
\address{Department of Physics and Engineering Physics, \\
Stevens Institute of Technology, Hoboken, NJ 07030 }
\author{Anatoly Yu. Smirnov}
\address{D-Wave Systems Inc., 320-1985 W. Broadway \\
Vancouver, British Columbia Canada V6J 4Y3}
\date{\today}
\maketitle

\begin{abstract}
{A fully microscopic theory of electron spin relaxation by the
D'yakonov-Perel' type spin-orbit coupling is developed for a semiconductor
quantum well in an ambient magnetic field applied perpendicular to the plane
of the well. We derive Bloch equations for an electron spin in the well and
determine explicit microscopic expressions for the spin relaxation times.
The dependencies of the electron spin relaxation rates on magnetic field
strength, temperature, and the lowest subband energy of the quantum well are
analyzed.}
\end{abstract}


\narrowtext

\section{Introduction}

Spin phenomenology in semiconductor structures has been at the focal point
of research interest over the past few years in connection with proposals
for spin-based quantum devices \cite{Awsh}. Spin implementation of quantum
computation, optical switches, magnetic memory cells, etc., calls for a
precise knowledge of spin dynamics and, in particular, the spin relaxation
rates. The basic mechanisms responsible for spin relaxation are those of
D'yakonov-Perel' (DP) \cite{DP,DK}, Elliott-Yafet \cite{EY}, and
Bir-Aronov-Pikus \cite{BAP}. It was shown \cite{Exp} that for III-V and
II-VI compounds, which are the most promising materials for device purposes,
the DP mechanism dominates at moderate temperatures and low hole
concentrations. The spin relaxation time due to the DP mechanism has
generally been expressed in the following semiphenomenological form \cite
{DP,DK}: 
\begin{equation}
\frac{1}{\tau _{s}}=Q\frac{\alpha ^{2}}{\hbar ^{2}\varepsilon _{g}}\tau
_{p}T^{3}  \label{DP}
\end{equation}
for bulk semiconductors, and 
\begin{equation}
\frac{1}{\tau _{s}}=\frac{\alpha ^{2}\left\langle p_{z}^{2}\right\rangle ^{2}%
}{2\hbar ^{2}m^{2}\varepsilon _{g}}\tau _{p}T  \label{DK}
\end{equation}
for quantum well structures, where $\alpha $ describes conduction band spin
splitting due to lack of inversion symmetry ($\alpha =0.07$ for GaAs), $%
\varepsilon _{g}$ is the band gap, and $T$ is the Kelvin temperature ($%
k_{B}=1$). The numerical coefficient $Q$ depends on the orbital scattering
mechanism, and $\left\langle p_{z}^{2}\right\rangle $ is the average square
of momentum in the quantum well growth direction. It should be noted that
these formulas involve the average momentum relaxation time, $\tau _{p}$, as
a phenomenological parameter.

In the present paper we develop a {\it fully microscopic} theory of spin
dynamics and apply it to a quantum well structure in the presence of an
external magnetic field, directed along the well growth direction, taking
account of various scattering mechanisms. Our theory facilitates the
microscopic determination of $\tau _{p}$ and its temperature and magnetic
field dependencies in terms of the fundamental material parameters.

Our analysis involves explicit recognition of the two stages in the
relaxation process corresponding to the relaxation time hierarchies involved
in (a) electron thermalization due to interaction with phonons, and (b) spin
relaxation. In the first stage of solution, (a), we determine the relaxation
rates and fluctuation characteristics of electron orbital motion due to
coupling to the phonon bath. Spin relaxation dynamics (the slowest process
in the system) can be neglected in this stage.

The second stage, (b), proceeds with analysis of the spin relaxation process
due to spin-orbit interaction, wherein the orbital degrees of freedom are
considered as an effective heat bath, having the characteristics determined
in the first stage. A standard set of Bloch equations with two distinct
relaxation times (longitudinal relaxation time, $T_{1}$, responsible for
spin magnetic moment relaxation, and transverse relaxation time, $T_{2}$,
responsible for decoherence) is derived in this second stage. In both stages
of our analyses we employ the method proposed in Ref. \cite{TOQS} and
developed in Ref. \cite{Mourokh,Spin}.

\section{Formulation}

The Hamiltonian of part (a) is given by 
\begin{equation}
H=H_{0}+U({\bf r}_{\perp },t)+H_{ph},
\end{equation}
where spin and its interactions are neglected in the first stage, and 
\begin{equation}
H_{0}=\frac{\hbar \omega _{0}}{2}+\frac{mV_{x}^{2}}{2}+\frac{mV_{y}^{2}}{2},
\end{equation}
is the Hamiltonian of a two-dimensional electron in a quantum well with
harmonic confinement in $z$-direction having frequency $\omega _{0}$. We
assume the temperature to be low enough, so that only the lowest energy
subband of the quantum well is occupied. In this case the motion of an
electron can be described by means of two electron in-plane velocity
components, which in the presence of a constant, uniform magnetic field
directed along the z-axis ${\bf B}=(0,0,B)$ are given by 
\begin{equation}
V_{x}=\frac{1}{m}\left( p_{x}-\frac{e}{c}A_{x}\right) ,\text{ }V_{y}=\frac{1%
}{m}\left( p_{y}-\frac{e}{c}A_{y}\right) ,\text{ }\left[ V_{x},V_{y}\right]
_{-}=-\frac{i\hbar \omega _{c}}{m},
\end{equation}
where ${\bf A}=(A_{x},A_{y},0)$ is the vector potential, $B=\frac{\partial }{%
\partial x}A_{y}-\frac{\partial }{\partial y}A_{x}$, $\omega _{c}=\left|
e\right| B/mc$ is the cyclotron frequency, and $\left[ ...,...\right] _{-}$
denotes a commutator. The phonon Hamiltonian has the form 
\begin{equation}
H_{ph}=\sum_{{\bf k}}\hbar \omega _{{\bf k}}(b_{{\bf k}}^{+}b_{{\bf k}}+{%
\frac{1}{2}}),
\end{equation}
where $\hbar \omega _{{\bf k}}$ is the phonon energy and $b_{{\bf k}}^{+}$
and $b_{{\bf k}}$ are creation and annihilation operators, respectively. The
term $U({\bf r}_{\perp },t)$ in Eq.(3) describes the electron coupling to
phonons and impurities.

The electron-phonon interaction is given by 
\begin{equation}
U_{e-ph}({\bf r}_{\perp },t)=-\sum_{{\bf k}}Q_{{\bf k}}(t)X_{{\bf k}}(t)=-%
\frac{1}{L^{3/2}}\sum_{{\bf k}}Q_{{\bf k}}(t)f(k_{z})e^{i{\bf k}_{\perp }%
{\bf r}_{\perp }},
\end{equation}
where 
\begin{equation}
Q_{{\bf k}}(t)=i\zeta (b_{{\bf k}}(t)-b_{-{\bf k}}^{+}(t))
\end{equation}
is the phonon heat bath variable ($\zeta $ is the strength of the
electron-phonon coupling), and the electron variable conjugate to this
operator is defined as 
\begin{equation}
X_{{\bf k}}(t)=\frac{1}{L^{3/2}}f(k_{z})e^{i{\bf k}_{\perp }{\bf r}_{\perp
}(t)}.
\end{equation}
Here, $L^{3}$ is the volume of the crystal, ${\bf r}_{\perp}=(x,y)$, and $%
f(k_{z})=\exp \left( -\frac{\hbar k_{z}^{2}}{4m\omega _{0}}\right) $ is the
electron confinement form factor. The response function of uncoupled phonon
heat bath variables, $\varphi _{{\bf k}}(t;t_{1}),$ and their correlation
function, $M_{{\bf k}}(t;t_{1}),$ are given by

\begin{equation}
\varphi _{{\bf k}}(t;t_{1})=\left\langle \frac{i}{\hbar }\left[ Q_{{\bf k}%
}^{(0)}(t),Q_{-{\bf k}}^{(0)}(t_{1})\right] _{-}\right\rangle \Theta
(t-t_{1})
\end{equation}
and

\begin{equation}
M_{{\bf k}}(t;t_{1})=\left\langle \frac{1}{2}\left[ Q_{{\bf k}}^{(0)}(t),Q_{-%
{\bf k}}^{(0)}(t_{1})\right] _{+}\right\rangle ,
\end{equation}
where $Q_{{\bf k}}^{(0)}(t)$ are the {\it unperturbed }phonon variables, $%
\Theta (t-t_{1})$ is the Heavyside unit step function, and $\left[ ...,...%
\right] _{+}$ denotes an anticommutator.

The electron-impurity interaction is described by 
\begin{equation}
U_{e-i}({\bf r}_{\perp })=-\frac{1}{L^{3/2}}\sum_{{\bf k}}U_{{\bf k}%
}f(k_{z})e^{i{\bf k}_{\perp }{\bf r}_{\perp }},
\end{equation}
where $U_{{\bf k}}$ are the spatial Fourier components of the impurity
potential with correlation function

\begin{equation}
\Phi _{{\bf k}}=\left\langle \frac{1}{2}\left[ U_{{\bf k}},U_{-{\bf k}}%
\right] _{+}\right\rangle .
\end{equation}

In summary, 
\begin{equation}
U({\bf r},t)=U_{e-ph}({\bf r}_{\perp },t)+U_{e-i}({\bf r}_{\perp })=-\sum_{%
{\bf k}}\left( Q_{{\bf k}}(t)+U_{{\bf k}}\right) X_{{\bf k}}(t).
\end{equation}

\section{Orbital Dynamics}

Employing the Hamiltonian of Eq.(3), we obtain Langevin-like operator
equations of motion determining in-plane electron dynamics, as given by (the
derivation is presented in Appendix A):

\begin{eqnarray}
\frac{d}{dt}V_{x}(t)+\omega _{c}V_{y}(t)+G_{x}\left[ V_{x}(t);V_{y}(t)\right]
&=&\xi _{x}(t),  \nonumber \\
\frac{d}{dt}V_{y}(t)-\omega _{c}V_{x}(t)+G_{y}\left[ V_{x}(t);V_{y}(t)\right]
&=&\xi _{y}(t),  \label{1}
\end{eqnarray}
with fluctuation sources $\xi _{x,y}(t)$ and collision terms $G_{x}\left[
V_{x}(t);V_{y}(t)\right] $ given by Eqs.(A8) and (A9). It should be
emphasized that the expressions for fluctuation sources (Eq.(A8)) are
obtained from microscopic analysis, and it is possible to calculate their
correlation functions of any order. In particular, the correlation function
of the fluctuation sources (in the case of weak coupling or Gaussian
statistics of the unperturbed phonon variables) is given by 
\begin{eqnarray}
K_{\alpha \beta }(t,t_{1}) &=&\left\langle \frac{1}{2}\left[ \xi _{\alpha
}(t),\xi _{\beta }(t_{1})\right] _{+}\right\rangle =\frac{1}{m^{2}}\sum_{%
{\bf k}}k_{\alpha }k_{\beta }f^{2}(k_{z})  \nonumber \\
&&\times \left( \left( M_{{\bf k}}(t,t_{1})+\Phi _{{\bf k}}\right)
\left\langle \frac{1}{2}\left[ X_{{\bf k}}(t),X_{-{\bf k}}(t_{1})\right]
_{+}\right\rangle +R_{{\bf k}}(\tau )\left\langle \frac{1}{2}\left[ X_{{\bf k%
}}(t),X_{-{\bf k}}(t_{1})\right] _{-}\right\rangle \right) ,
\end{eqnarray}
where $R_{{\bf k}}(t-t_{1})=\hbar \left( \varphi _{{\bf k}}(t-t_{1})+\varphi
_{{\bf k}}(t_{1}-t)\right) /2i$ .

We can separate the electron velocity operator into its average and
fluctuation parts, 
\begin{equation}
V_{x,y}(t)=\left\langle V_{x,y}(t)\right\rangle +\widetilde{V}_{x,y}(t).
\end{equation}
Due to relaxation processes only the fluctuating part is nonzero, $%
\left\langle V_{x,y}(t)\right\rangle =0$, and the equations of motion for
the fluctuating components take the forms 
\begin{eqnarray}
\frac{d}{dt}\widetilde{V}_{x}(t)+\omega _{c}\widetilde{V}_{y}(t)+G_{x}\left[
V_{x}(t);V_{y}(t)\right] -\left\langle G_{x}\left[ V_{x}(t);V_{y}(t)\right]
\right\rangle &=&\xi _{x}(t)  \nonumber \\
\frac{d}{dt}\widetilde{V}_{y}(t)-\omega _{c}\widetilde{V}_{x}(t)+G_{y}\left[
V_{x}(t);V_{y}(t)\right] -\left\langle G_{y}\left[ V_{x}(t);V_{y}(t)\right]
\right\rangle &=&\xi _{y}(t).  \label{4}
\end{eqnarray}

To further simplify Eqs. (18), we have to calculate the (anti)commutators
involved in Eqs. (16), (A8), and (A9). This procedure is presented in
Appendix B resulting in the following simplified equations: 
\begin{eqnarray}
\left( \frac{d}{dt}+\gamma _{0}\right) \widetilde{V}_{x}(t)+\left( \omega
_{c}+\delta \right) \widetilde{V}_{y}(t) &=&\xi _{x}(t),  \nonumber \\
\left( \frac{d}{dt}+\gamma _{0}\right) \widetilde{V}_{y}(t)-\left( \omega
_{c}-\delta \right) \widetilde{V}_{x}(t) &=&\xi _{y}(t),  \label{Velocity}
\end{eqnarray}
where the damping rate, $\gamma _{0}=\gamma _{x}=\gamma _{y}$, and the
frequency shift, $\delta =\delta _{x}=\delta _{y}$, are given by 
\begin{eqnarray}
\gamma _{x,y} &=&\frac{1}{mL^{3}}\sum_{{\bf k}}k_{x,y}^{2}f^{2}(k_{z})%
\int_{0}^{+\infty }d\tau \tau \left\{ \left( M_{{\bf k}}(\tau )+\Phi _{{\bf k%
}}\right) \frac{2}{\hbar }\sin \left( \frac{\hbar k_{\perp }^{2}}{2m}\tau
\right) \right.  \nonumber \\
&&\left. +\varphi _{{\bf k}}(\tau )\cos \left( \frac{\hbar k_{\perp }^{2}}{2m%
}\tau \right) \right\} \exp \left\{ -\frac{\tau ^{2}}{2\tau _{c}^{2}({\bf k}%
_{\perp })}\right\}  \label{5}
\end{eqnarray}
and 
\begin{eqnarray}
\delta _{x,y} &=&\frac{1}{mL^{3}}\sum_{{\bf k}}k_{x}k_{y}f^{2}(k_{z})%
\int_{0}^{+\infty }d\tau \tau \left\{ \left( M_{{\bf k}}(\tau )+\Phi _{{\bf k%
}}\right) \frac{2}{\hbar }\sin \left( \frac{\hbar k_{\perp }^{2}}{2m}\tau
\right) \right.  \nonumber \\
&&\left. +\varphi _{{\bf k}}(\tau )\cos \left( \frac{\hbar k_{\perp }^{2}}{2m%
}\tau \right) \right\} \exp \left\{ -\frac{\tau ^{2}}{2\tau _{c}^{2}({\bf k}%
_{\perp })}\right\} .
\end{eqnarray}
Here, $\tau =t-t_{1}$ and $\tau _{c}^{2}({\bf k}_{\perp })$ is given by
Eq.(B11). It follows from Eq. (\ref{Velocity}) that the Fourier transforms
of velocity correlation functions are given by 
\begin{equation}
\left\langle \frac{1}{2}\left[ V_{\alpha }(\omega );V_{\beta }\right]
_{+}\right\rangle =\int d\tau e^{i\omega \tau }\left\langle \frac{1}{2}\left[
V_{\alpha }(t+\tau ),V_{\beta }(t)\right] _{+}\right\rangle ,\text{ }\alpha
,\beta =x,y,
\end{equation}
\begin{equation}
\left\langle \frac{1}{2}\left[ V_{x}(\omega );V_{x}\right] _{+}\right\rangle
=\left\langle \frac{1}{2}\left[ V_{y}(\omega );V_{y}\right]
_{+}\right\rangle =\frac{K(\omega )}{2}\left( \frac{1}{\left( \omega -\omega
_{c}\right) ^{2}+\gamma _{0}^{2}}+\frac{1}{\left( \omega +\omega _{c}\right)
^{2}+\gamma _{0}^{2}}\right) ,
\end{equation}
and 
\begin{equation}
\left\langle \frac{1}{2}\left[ V_{x}(\omega );V_{y}\right] _{+}\right\rangle
=-\left\langle \frac{1}{2}\left[ V_{y}(\omega );V_{x}\right]
_{+}\right\rangle =\frac{K(\omega )}{2i}\left( \frac{1}{\left( \omega
-\omega _{c}\right) ^{2}+\gamma _{0}^{2}}-\frac{1}{\left( \omega +\omega
_{c}\right) ^{2}+\gamma _{0}^{2}}\right) ,
\end{equation}
where we neglected terms of the of order $\delta /\omega _{c}<<1$. The
correlation function of the fluctuation forces is defined as $K(\omega
)=K_{\alpha \alpha }(\omega )$, where 
\begin{eqnarray}
K_{\alpha \beta }(\omega ) &=&\int dt\tau e^{i\omega \tau }K_{\alpha \beta
}(t+\tau ,t)  \nonumber \\
&=&\frac{1}{m^{2}}\int \frac{d^{3}{\bf k}}{\left( 2\pi \right) ^{3}}%
k_{\alpha }k_{\beta }f^{2}(k_{z})\int_{0}^{+\infty }d\tau \tau \left\{
\left( M_{{\bf k}}(\tau )+\Phi _{{\bf k}}\right) \sin \left( \frac{\hbar
k_{\perp }^{2}}{2m}\tau \right) -\right.  \nonumber \\
&&\left. -iR_{{\bf k}}(\tau )\cos \left( \frac{\hbar k_{\perp }^{2}}{2m}\tau
\right) \right\} \exp \left\{ -\frac{\tau ^{2}}{2\tau _{c}^{2}({\bf k}%
_{\perp })}\right\} .
\end{eqnarray}

To carry out the $\tau -$ and ${\bf k}-$integrations in Eqs.(20) and (25),
we have to employ explicit expressions for the response and correlation
functions for a particular scattering mechanism. In the present paper we
consider polar optical phonons, deformational acoustic phonons and charged
impurities as possible scattering mechanisms. Their corresponding response
and correlation functions are listed below.

Optical phonons: 
\begin{equation}
\varphi _{{\bf k}}^{OP}(\tau )=\frac{4\pi \Omega _{0}e^{2}}{k^{2}\epsilon
^{\ast }}\sin \left( \Omega _{0}\tau \right) \eta (\tau ),\text{ }M_{{\bf k}%
}^{OP}(\tau )=\frac{\hbar }{2}\frac{4\pi \Omega _{0}e^{2}}{k^{2}\epsilon
^{\ast }}\cos \left( \Omega _{0}\tau \right) \coth (\frac{\hbar \Omega _{0}}{%
2k_{B}T}),
\end{equation}
where $\Omega _{0}$ is the optical phonon frequency, $1/\epsilon ^{\ast
}=1/\epsilon _{\infty }-1/\epsilon _{0}$, ($\epsilon _{\infty }$ and $%
\epsilon _{0}$ are the hf and static permittivities of the crystal,
respectively).

Deformational acoustic phonons: 
\begin{equation}
\varphi _{{\bf k}}^{AP}(\tau )=\frac{D^{2}k}{\rho u}\sin \left( uk\tau
\right) \eta (\tau ),\text{ }M_{{\bf k}}^{AP}(\tau )=\frac{\hbar }{2}\frac{%
D^{2}k}{\rho u}\cos \left( uk\tau \right) \coth (\frac{\hbar uk}{2k_{B}T}),
\end{equation}
where $D$ is the deformation potential , $\rho $ is the crystal density, and 
$u$ is the sound velocity.

Static charged impurities have only correlation function, given by 
\begin{equation}
\Phi _{{\bf k}}=\frac{2e^{4}n_{t}^{\ast }}{\pi \epsilon _{0}^{2}\left(
k^{2}+r_{0}^{-2}\right) ^{2}},
\end{equation}
where $r_{0}$ is the screening radius, $n_{t}^{\ast }=\sum_{\alpha
}n_{\alpha }Z_{a}^{2}$, $n_{\alpha }$ is the impurity concentration for
species $\alpha $, and $Z_{\alpha }$ is their charge number.

We neglect the cross correlations of scattering processes and, consequently,
they are additive in both the damping rate and in the correlation function
of the fluctuation sources: 
\begin{eqnarray}
\gamma _{0} &=&\gamma _{0}^{OP}+\gamma _{0}^{I}+\gamma _{0}^{AP},  \nonumber
\\
\widetilde{K}(\omega ) &=&\widetilde{K}^{OP}(\omega )+\widetilde{K}%
^{I}(\omega )+\widetilde{K}^{AP}(\omega ).  \label{gammaK}
\end{eqnarray}
Substituting the corresponding response and correlation functions in
Eqs.(20) and (25), we obtain the contributions of polar optic phonons as 
\begin{eqnarray}
\gamma _{0}^{OP} &=&\frac{1}{\sqrt{2\pi }}\frac{\Omega _{0}e^{2}}{m\epsilon
^{\ast }}\int_{0}^{+\infty }dk_{\perp }k_{\perp }^{3}\int_{0}^{+\infty
}dk_{z}\frac{\tau _{c}^{3}(k_{\perp })f^{2}(k_{z})}{k^{2}}  \nonumber \\
&&\left\{ \left( N_{0}+1\right) \left( \omega _{\perp }+\Omega _{0}\right)
\exp \left[ -\frac{1}{2}\left( \omega _{\perp }+\Omega _{0}\right) ^{2}\tau
_{c}^{2}(k_{\perp })\right] +\right.  \nonumber \\
&&\left. +N_{0}\left( \omega _{\perp }-\Omega _{0}\right) \exp \left[ -\frac{%
1}{2}\left( \omega _{\perp }-\Omega _{0}\right) ^{2}\tau _{c}^{2}(k_{\perp })%
\right] \right\}  \label{OP1}
\end{eqnarray}
and 
\begin{eqnarray}
\widetilde{K}^{OP}(\omega ) &=&\frac{1}{2\sqrt{2\pi }}\frac{\hbar \Omega
_{0}e^{2}}{m^{2}\epsilon ^{\ast }}\int_{0}^{+\infty }dk_{\perp }k_{\perp
}^{3}\int_{0}^{+\infty }dk_{z}\frac{\tau _{c}(k_{\perp })f^{2}(k_{z})}{k^{2}}
\nonumber \\
&&\left\{ \left( N_{0}+1\right) \left( \exp \left[ -\frac{1}{2}\left( \omega
+\omega _{\perp }+\Omega _{0}\right) ^{2}\tau _{c}^{2}(k_{\perp })\right]
+\right. \right.  \nonumber \\
&&\left. +\exp \left[ -\frac{1}{2}\left( \omega -\omega _{\perp }-\Omega
_{0}\right) ^{2}\tau _{c}^{2}(k_{\perp })\right] \right) +  \nonumber \\
&&+N_{0}\left( \exp \left[ -\frac{1}{2}\left( \omega +\omega _{\perp
}-\Omega _{0}\right) ^{2}\tau _{c}^{2}(k_{\perp })\right] +\right.  \nonumber
\\
&&\left. \left. +\exp \left[ -\frac{1}{2}\left( \omega -\omega _{\perp
}+\Omega _{0}\right) ^{2}\tau _{c}^{2}(k_{\perp })\right] \right) \right\} ;
\label{OP2}
\end{eqnarray}
the contributions of deformational acoustic phonons as 
\begin{eqnarray}
\gamma _{0}^{AP} &=&\frac{1}{4\pi \sqrt{2\pi }}\frac{D^{2}}{m\rho u}%
\int_{0}^{+\infty }dk_{\perp }k_{\perp }^{3}\int_{0}^{+\infty }dk_{z}k\tau
_{c}^{3}(k_{\perp })f^{2}(k_{z})  \nonumber \\
&&\left\{ \left( N_{k}+1\right) \left( \omega _{\perp }+uk\right) \exp \left[
-\frac{1}{2}\left( \omega _{\perp }+uk\right) ^{2}\tau _{c}^{2}(k_{\perp })%
\right] +\right.  \nonumber \\
&&\left. +N_{k}\left( \omega _{\perp }-uk\right) \exp \left[ -\frac{1}{2}%
\left( \omega _{\perp }-uk\right) ^{2}\tau _{c}^{2}(k_{\perp })\right]
\right\}  \label{AP1}
\end{eqnarray}
and 
\begin{eqnarray}
\widetilde{K}^{AP}(\omega ) &=&\frac{1}{8\pi \sqrt{2\pi }}\frac{\hbar D^{2}}{%
m^{2}\rho u}\int_{0}^{+\infty }dk_{\perp }k_{\perp }^{3}\int_{0}^{+\infty
}dk_{z}k\tau _{c}(k_{\perp })f^{2}(k_{z})  \nonumber \\
&&\left\{ \left( N_{k}+1\right) \left( \exp \left[ -\frac{1}{2}\left( \omega
+\omega _{\perp }+uk\right) ^{2}\tau _{c}^{2}(k_{\perp })\right] +\right.
\right.  \nonumber \\
&&\left. +\exp \left[ -\frac{1}{2}\left( \omega -\omega _{\perp }-uk\right)
^{2}\tau _{c}^{2}(k_{\perp })\right] \right) +  \nonumber \\
&&+N_{k}\left( \exp \left[ -\frac{1}{2}\left( \omega +\omega _{\perp
}-uk\right) ^{2}\tau _{c}^{2}(k_{\perp })\right] +\right.  \nonumber \\
&&\left. \left. +\exp \left[ -\frac{1}{2}\left( \omega -\omega _{\perp
}+uk\right) ^{2}\tau _{c}^{2}(k_{\perp })\right] \right) \right\} ;
\label{AP2}
\end{eqnarray}
and, finally, the contributions of the charged impurities as 
\begin{equation}
\gamma _{0}^{I}=\frac{1}{2\pi ^{2}\sqrt{2\pi }}\frac{e^{4}n_{t}^{\ast }}{%
m^{2}\epsilon _{0}^{2}}\int_{0}^{+\infty }dk_{\perp }k_{\perp
}^{5}\int_{0}^{+\infty }dk_{z}\frac{\tau _{c}^{3}(k_{\perp })f^{2}(k_{z})}{%
\left( k^{2}+r_{0}^{-2}\right) ^{2}}\exp \left( -\frac{\omega _{\perp
}^{2}\tau _{c}^{2}(k_{\perp })}{2}\right) ,  \label{I1}
\end{equation}
and 
\begin{eqnarray}
\widetilde{K}^{I}(\omega ) &=&\frac{1}{2\pi ^{2}\sqrt{2\pi }}\frac{%
e^{4}n_{t}^{\ast }}{m^{2}\epsilon _{0}^{2}}\int_{0}^{+\infty }dk_{\perp
}k_{\perp }^{3}\int_{0}^{+\infty }dk_{z}\frac{\tau _{c}(k_{\perp
})f^{2}(k_{z})}{\left( k^{2}+r_{0}^{-2}\right) ^{2}}\cdot  \nonumber \\
&&\cdot \left( \exp \left[ -\frac{1}{2}\left( \omega +\omega _{\perp
}\right) ^{2}\tau _{c}^{2}(k_{\perp })\right] +\exp \left[ -\frac{1}{2}%
\left( \omega -\omega _{\perp }\right) ^{2}\tau _{c}^{2}(k_{\perp })\right]
\right) .  \label{I2}
\end{eqnarray}
In all these formulae we use the notation $k=\sqrt{k_{\perp }^{2}+k_{z}^{2}} 
$ and $\omega _{\perp }=\hbar k_{\perp }^{2}/2m$. $N_{0}=\left[ \exp \left(
-\hbar \Omega _{0}/T\right) -1\right] ^{-1}$ and $N_{k}=\left[ \exp \left(
-\hbar uk/T\right) -1\right] ^{-1}$ are Bose distribution functions for the
optical and acoustic phonons, respectively. The level of thermal velocity
fluctuations, $\left\langle V_{x}^{2}\right\rangle $, can be determined
self-consistently using $\left\langle V_{x}^{2}\right\rangle =\widetilde{K}%
(\omega _{c})/2\gamma _{0}$ \cite{Mourokh}.

\section{Spin Dynamics}

In the second stage we analyze spin relaxation due to DP interaction between
spin and electron orbital motion. The corresponding spin interaction
Hamiltonian is given by \cite{DP,DK} 
\begin{equation}
H_{spin}=\mu_B\left( \overrightarrow{\sigma }\cdot \overrightarrow{B}\right)
+ \frac{\hbar }{2}\left( \overrightarrow{\sigma }\cdot \overrightarrow{%
\Omega }\right) ,
\end{equation}
where $\mu_B\left( \overrightarrow{\sigma }\cdot \overrightarrow{B}\right) $
is the Zeeman splitting Hamiltonian term, $\mu_B$ is the Bohr magneton, and 
\begin{equation}
\overrightarrow{\Omega }=\frac{\alpha \overrightarrow{\varkappa }}{\hbar
m^{3/2}\sqrt{2\varepsilon _{g}}},
\end{equation}
with 
\begin{equation}
\varkappa _{x}=m^{3}V_{x}\left( V_{y}^{2}-V_{z}^{2}\right) ,\text{ }%
\varkappa _{y}=m^{3}V_{y}\left( V_{z}^{2}-V_{x}^{2}\right) ,\text{ }%
\varkappa _{z}=m^{3}V_{z}\left( V_{x}^{2}-V_{y}^{2}\right) .
\end{equation}
For a quantum well strongly confined in $<001>$ direction, $H_{DP}$ can be
simplified as \cite{DK} 
\begin{equation}
H_{DP}=-\sigma _{x}Q_{x}(t)-\sigma _{y}Q_{y}(t),
\end{equation}
where 
\begin{equation}
Q_{x}(t)=\lambda V_{x}(t),\text{ }Q_{y}(t)=-\lambda V_{y}(t),
\end{equation}
and 
\begin{equation}
\lambda =\frac{\alpha \left\langle p_{z}^{2}\right\rangle }{2\sqrt{%
2m\varepsilon _{g}}},
\end{equation}
with $\left\langle p_{z}^{2}\right\rangle =m\hbar \omega _{0}/2$ if the
confinement is parabolic.

Considering orbital motion to play the role of an effective heat bath with
correlation functions given by Eqs.(23) and (24), and employing a second
application of the method of Refs. \cite{TOQS,Mourokh,Spin}, we obtain a set
of Bloch equations for the average spin projections as 
\begin{eqnarray}
\frac{d}{dt}\left\langle \sigma _{x}(t)\right\rangle  &=&-\frac{\left\langle
\sigma _{x}(t)\right\rangle }{T_{2}}-\left( \omega _{B}+\delta _{x}\right)
\left\langle \sigma _{y}(t)\right\rangle ,  \nonumber \\
\frac{d}{dt}\left\langle \sigma _{y}(t)\right\rangle  &=&\left( \omega
_{B}+\delta _{y}\right) \left\langle \sigma _{x}(t)\right\rangle -\frac{%
\left\langle \sigma _{y}(t)\right\rangle }{T_{2}},  \nonumber \\
\frac{d}{dt}\left\langle \sigma _{z}(t)\right\rangle  &=&\frac{\sigma
_{z}^{0}-\left\langle \sigma _{z}(t)\right\rangle }{T_{1}},  \label{Bloch}
\end{eqnarray}
where $\sigma _{x},\sigma _{y},\sigma _{z}$ are the Pauli matrices and $%
\sigma _{z}^{0}=-\tanh \left( \hbar \omega _{B}/2T\right) $ is the
equilibrium $z$-component of spin. $\omega _{B}=g\mu _{B}B/\hbar $, $\mu
_{B}=\left| e\right| \hbar /2m_{0}c$ is the Bohr magneton, and the $g$%
-factor is $-0.44$ for GaAs. It should be emphasized that we determine the
relaxation times $T_{1}$ and $T_{2}$ microscopically as 
\begin{equation}
T_{1}=\frac{\tau _{s}}{2},\text{ }T_{2}=\tau _{s},
\end{equation}
where 
\begin{equation}
\frac{1}{\tau _{s}}=\frac{2\lambda ^{2}}{\hbar ^{2}}\frac{\widetilde{K}%
(\omega _{B})}{\left( \omega _{B}-\omega _{c}\right) ^{2}+\gamma _{0}^{2}}.
\label{rate}
\end{equation}
While this result is reminiscent of a formula obtained by Ivchenko for bulk
semiconductors \cite{Ivchenko}, that differs from the present quantum well
case under consideration here in that the Ivchenko result has a sum of
several Lorentzians and involves numerical constants in place of our
relaxation rate $\gamma _{0}$ and numerator function $\widetilde{K}(\omega )$
which have magnetic field and temperature dependencies that are explicitly
determined by Eq. (\ref{gammaK}) on the microscopic basis. It is evident
from Eq. (\ref{rate}), that the effect of magnetic field on spin relaxation
is twofold: Firstly, there is magnetic field dependence of the fluctuation
source correlator, $\widetilde{K}(\omega _{B})$, associated with electron
transitions between energy levels having different spin. Secondly, the
difference between the frequencies $\omega _{B}$ and $\omega _{c}$ is
involved in the denominator of Eq. (\ref{rate}), which occurs because of
deviations of electron effective mass and g-factor in semiconductors from
the free electron values. The temperature dependence can also be separated
into two parts, (i) via electron thermal fluctuations (this contribution is
linear with increasing temperature); and (ii) via temperature dependence of
the momentum relaxation rate, $\gamma _{0}=\gamma _{0}(T)$.

In the absence of a magnetic field ($\omega _{c}=\omega _{B}=0$, $%
\left\langle V_{x}^{2}\right\rangle =\widetilde{K}(0)/2\gamma _{0}=T/m$), we
recover Eq. (\ref{DK}) as the zero field limit: 
\begin{equation}
\left. \frac{1}{\tau _{s}}\right| _{B=0}=\frac{4\lambda ^{2}}{\hbar ^{2}}%
\frac{1}{\gamma _{0}}\frac{\widetilde{K}(0)}{2\gamma _{0}}.=\frac{4\lambda
^{2}}{\hbar ^{2}}\frac{1}{\gamma _{0}}\left\langle V_{x}^{2}\right\rangle =%
\frac{4\lambda ^{2}}{\hbar ^{2}}\frac{1}{\gamma _{0}}\frac{T}{m}=\frac{%
\alpha ^{2}\left\langle p_{z}^{2}\right\rangle ^{2}}{2\hbar
^{2}m^{2}\varepsilon _{g}}\frac{1}{\gamma _{0}}T,
\end{equation}
wherein our microscopic analysis yields the phenomenological constant $\tau
_{p}$ of Eqs. (\ref{DP},\ref{DK}) as $\tau _{p}\rightarrow $ $1/\gamma _{0}$.

Figs. 1(a,b) exhibit the dependence of the relaxation rate on the energy of
the lowest electronic subband of the quantum well (which is given by $%
E_{0}=\hbar \omega _{0}/2$ for the case of harmonic confinement) for various
temperatures and magnetic field strengths. It is evident from these figures
that at low temperature the applied magnetic field suppresses spin
relaxation, whereas at higher temperatures the momentum relaxation rate
dominates in the denominator of Eq. (\ref{rate}) and there is negligible
magnetic field dependence in this case. In these calculations we have
employed the following set of parameters for a GaAs-based quantum well:
electron effective mass $m=0.067m_{0}$, $1/\epsilon ^{\ast }=0.0069,$
optical phonon energy $\hbar \Omega _{0}=0.035$ $eV$, deformation potential $%
D=8.6$ $eV$, density $\rho =5.4$ $g$ $cm^{-3}$, sound velocity $u=5\cdot
10^{5}$ $cm$ $s^{-1}$, impurity concentration $n_{t}^{\ast }=10^{17}$ $%
cm^{-3}$, and static permittivity $\epsilon _{0}=13$. It should be
emphasized that our microscopic calculations yield {\it quantitative}
agreement with the experimental results of Ref. \cite{Exp}. The full
magnetic field dependence (up to $1000$ $Gs$) of the spin relaxation rate is
presented in Fig. 2 for $\hbar \omega _{0}=0.01$ $eV$. This dependence has
the well-known \cite{DP} Lorentzian shape, evident in Eq. (\ref{rate}), and
for a reasonable range of magnetic field strength it is not affected by the
magnetic field dependence of the velocity fluctuations, which are embedded
in the function $\widetilde{K}(\omega _{B})$. (However, for the case of
dilute magnetic semiconductors having a large electron $g$-factor, to which
our general analysis is also applicable, the magnetic field dependence of
the function $\widetilde{K}(\omega _{B})$ can be of crucial importance.) The
magnetic field dependence for the case of bulk semiconductors (represented
by a sum of several Lorentzians) was obtained in Ref. \cite{Ivchenko} and is
also reconfirmed by our microscopic analysis \cite{Puller}.

The temperature dependence of the spin relaxation rate is shown in Fig. 3.
The non-monotonic behavior is due to the dominance of different scattering
mechanisms in different temperature ranges. At low temperatures (before the
peak), the momentum scattering rate is determined by impurities and is
almost independent of temperature, and, consequently, we have almost linear
growth of the spin relaxation rate with increasing temperature, as predicted
by Eq. (\ref{DK}). However, as temperature further increases, optical
phonons become the dominant scattering mechanism and, with this, the
momentum scattering time becomes temperature dependent, resulting in
suppression of the relaxation rate. It is important to note that in the
temperature range in which phonons dominate, the relaxation rate is almost
independent of the magnetic field, whereas at low temperatures the magnetic
field shifts the peak and gives rise to deviations from linear behavior.

\section{Conclusions}

In conclusion, we have developed a fully microscopic theory of electron-spin
relaxation in semiconductors by the D'yakonov-Perel' mechanism. We have
applied this theory to a quantum well structure with a magnetic field in the
growth direction. A set of Bloch equations for a spin system has been
derived with microscopically determined longitudinal, $T_{1}$, and
transverse, $T_{2}$, relaxation times,\ which are related as $T_{1}=T_{2}/2$
for the quantum well structure grown in $<001>$ direction \cite{DK}. The
well-known semiphenomenological expression for the spin relaxation rate \cite
{DK} emerges as the zero-magnetic field limit of our result. Furthermore, we
have analyzed the dependencies of the electron-spin relaxation rate on the
energy of the lowest quantum well subband, on magnetic field strength, and
on temperature obtaining {\it quantitative} agreement with the experimental
results.

{\bf APPENDIX A}

The derivation of Langevin-like equations for electron velocity operators,
Eq.(15) presented here follows the approach developed in our previous works 
\cite{TOQS,Mourokh,Spin}. We start from Heisenberg equations of motion for
electron position and velocity operators for the Hamiltonian of Eq.(3), as 
\begin{eqnarray}
\frac{d}{dt}x(t) &=&\frac{1}{i\hbar }\left[ x(t),H\right] _{-}=V_{x}(t), 
\nonumber \\
\frac{d}{dt}y(t) &=&\frac{1}{i\hbar }\left[ y(t),H\right] _{-}=V_{y}(t), 
\nonumber \\
\frac{d}{dt}V_{x}(t) &=&\frac{1}{i\hbar }\left[ V_{x}(t),H\right]
_{-}=-\omega _{c}V_{y}(t)+\frac{1}{m}\sum_{{\bf k}}ik_{x}\left( Q_{{\bf k}%
}(t)+U_{{\bf k}}\right) X_{{\bf k}}(t),  \nonumber \\
\frac{d}{dt}V_{y}(t) &=&\frac{1}{i\hbar }\left[ V_{y}(t),H\right]
_{-}=\omega _{c}V_{x}(t)+\frac{1}{m}\sum_{{\bf k}}ik_{y}\left( Q_{{\bf k}%
}(t)+U_{{\bf k}}\right) X_{{\bf k}}(t).  \eqnum{A1}
\end{eqnarray}
If the coupling between the electron subsystem and heat bath is weak, or in
the case of Gaussian statistics of unperturbed heat bath variables, the
fully coupled Heisenberg heat bath variable is given by \cite{TOQS} 
\begin{equation}
Q_{{\bf k}}(t)=Q_{{\bf k}}^{(0)}(t)+\int_{-\infty }^{t}dt_{1}\varphi _{{\bf k%
}}(t;t_{1})X_{-{\bf k}}(t_{1}),  \eqnum{A2}
\end{equation}
where the response function of the phonon heat bath $\varphi _{{\bf k}%
}(t;t_{1})$ is defined by Eq.(10). Substituting Eq.(A2) into Eq.(A1), we
have to take into account the fact that only the fully coupled heat bath
variable commutes with the electron variable taken at equal times.
Accordingly, we perform a symmetrization of both terms of Eq.(A2) with the
electron variables, with the results 
\begin{eqnarray}
\frac{d}{dt}V_{x}(t)+\omega _{c}V_{y}(t) &=&\frac{1}{m}\sum_{{\bf k}%
}ik_{x}\left( \frac{1}{2}\left[ U_{{\bf k}},X_{{\bf k}}(t)\right] _{+}+\frac{%
1}{2}\left[ Q_{{\bf k}}^{(0)}(t),X_{{\bf k}}(t)\right] _{+}+\right.  
\nonumber \\
&&\left. +\int_{-\infty }^{t}dt_{1}\varphi _{{\bf k}}(t;t_{1})\frac{1}{2}%
\left[ X_{{\bf k}}(t),X_{-{\bf k}}(t_{1})\right] _{+}\right) ,  \nonumber \\
\frac{d}{dt}V_{y}(t)-\omega _{c}V_{x}(t) &=&\frac{1}{m}\sum_{{\bf k}%
}ik_{y}\left( \frac{1}{2}\left[ U_{{\bf k}},X_{{\bf k}}(t)\right] _{+}+\frac{%
1}{2}\left[ Q_{{\bf k}}^{(0)}(t),X_{{\bf k}}(t)\right] _{+}+\right.  
\nonumber \\
&&\left. \int_{-\infty }^{t}dt_{1}\varphi _{{\bf k}}(t;t_{1})\frac{1}{2}%
\left[ X_{{\bf k}}(t),X_{-{\bf k}}(t_{1})\right] _{+}\right) .  \eqnum{A3}
\end{eqnarray}
To eliminate the unperturbed phonon/impurity variables we employ the
Furutsu-Novikov theorem \cite{Furutsu}: 
\begin{equation}
\left\langle \frac{1}{2}\left[ Q_{{\bf k}}^{(0)}(t),X_{{\bf k}}(t)\right]
_{+}\right\rangle =\int_{-\infty }^{+\infty }dt_{1}M_{{\bf k}%
}(t;t_{1})\left\langle \frac{\delta X_{{\bf k}}(t)}{\delta Q_{-{\bf k}%
}^{(0)}(t_{1})}\right\rangle ,  \eqnum{A4}  \label{FN}
\end{equation}
where the correlation function $M_{{\bf k}}(t;t_{1})$ of free phonon
variables is given by Eq.(11) and $\delta /\delta Q^{(0)}(t_{1})$ is the
functional derivative with respect to the uncoupled heat bath variable $%
Q^{(0)}(t_{1})$. Eq.(A4) can be derived by an application of the Wick
theorem, assuming the operator $X_{{\bf k}}(t)$ to be a functional of $%
\{Q^{(0)}(t_{1})\}$ with $t_{1}\leq t$ (See Ref. \cite{TOQS}). This
expression is exact only for Gaussian statistics of the variables $Q^{(0)}(t)
$, and it also can be applied in the case of weak coupling. For the case of
strong coupling, Eq.(A4) requires modification to include functional
derivatives of all orders. The functional derivative on the right-hand side
of Eq. (\ref{FN}) is proportional to the commutator \cite{TOQS} in the form 
\begin{equation}
\left\langle \frac{\delta X_{{\bf k}}(t)}{\delta Q_{-{\bf k}}^{(0)}(t)}%
\right\rangle =\left\langle \frac{i}{\hbar }\left[ X_{{\bf k}}(t),X_{-{\bf k}%
}(t_{1})\right] _{-}\right\rangle \Theta (t-t_{1}),  \eqnum{A5}
\end{equation}
with the following result 
\begin{equation}
\left\langle \frac{1}{2}\left[ Q_{{\bf k}}^{(0)}(t),X_{{\bf k}}(t)\right]
_{+}\right\rangle =\int_{-\infty }^{t}dt_{1}M_{{\bf k}}(t;t_{1})\left\langle 
\frac{i}{\hbar }\left[ X_{{\bf k}}(t),X_{-{\bf k}}(t_{1})\right]
_{-}\right\rangle .  \eqnum{A6}
\end{equation}
An analogous treatment of the electron-impurity correlation term yields 
\begin{equation}
\left\langle \frac{1}{2}\left[ U_{{\bf k}}(t),X_{{\bf k}}(t)\right]
_{+}\right\rangle =\int_{-\infty }^{t}dt_{1}\Phi _{{\bf k}}\left\langle 
\frac{i}{\hbar }\left[ X_{{\bf k}}(t),X_{-{\bf k}}(t_{1})\right]
_{-}\right\rangle ,  \eqnum{A7}
\end{equation}
where $\Phi _{{\bf k}}$ is the impurity potential correlation function given
by Eq.(13). We also introduce fluctuation source operators defined as 
\begin{eqnarray}
\xi _{x,y}(t) &=&\frac{1}{m}\sum_{{\bf k}}ik_{x,y}\left( \frac{1}{2}\left[
Q_{{\bf k}}^{(0)}(t)+U_{{\bf k}},X_{{\bf k}}(t)\right] _{+}-\right.  
\nonumber \\
&&\left. -\int_{-\infty }^{t}dt_{1}\left( M_{{\bf k}}(t;t_{1})+\Phi _{{\bf k}%
}\right) \frac{i}{\hbar }\left[ X_{{\bf k}}(t),X_{-{\bf k}}(t_{1})\right]
_{-}\right)   \eqnum{A8}
\end{eqnarray}
with zero average, $\left\langle \xi _{x,y}(t)\right\rangle =0$, and
collision terms defined as 
\begin{eqnarray}
G_{x,y}\left[ V_{x}(t);V_{y}(t)\right]  &=&-\frac{1}{m}\sum_{{\bf k}%
}ik_{x,y}\int_{-\infty }^{t}dt_{1}\left( \left( M_{{\bf k}}(t;t_{1})+\Phi _{%
{\bf k}}\right) \frac{i}{\hbar }\left[ X_{{\bf k}}(t),X_{-{\bf k}}(t_{1})%
\right] _{-}+\right.   \nonumber \\
&&\left. +\varphi _{{\bf k}}(t;t_{1})\frac{1}{2}\left[ X_{{\bf k}}(t),X_{-%
{\bf k}}(t_{1})\right] _{+}\right) ,  \eqnum{A9}  \label{2}
\end{eqnarray}
resulting in the Langevin-like equations of Eq.(15).

\bigskip {\bf APPENDIX B}

In this appendix we determine the commutators $\left[ X_{{\bf k}}(t),X_{-%
{\bf k}}(t_{1})\right] _{\pm }$ involved in Eqs.(16), (A8), and (A9). We
assume that there is finite correlation time, $\tau _{c}$, of the
electron-phonon interaction (which will be obtained self-consistently). For
the case of weak coupling, the commutators can be calculated using the free,
uncoupled evolution of the electron velocity operators during $\tau _{c}$ ,
as given by 
\begin{eqnarray}
x(t) &=&x(t_{1})+V_{x}(t_{1})\frac{\sin (\omega _{c}(t-t_{1}))}{\omega _{c}}%
-V_{y}(t_{1})\frac{1-\cos (\omega _{c}(t-t_{1}))}{\omega _{c}},  \eqnum{B1}
\\
y(t) &=&y(t_{1})+V_{x}(t_{1})\frac{1-\cos (\omega _{c}(t-t_{1}))}{\omega _{c}%
}+V_{y}(t_{1})\frac{\sin (\omega _{c}(t-t_{1}))}{\omega _{c}}.  \nonumber
\end{eqnarray}
This allows to us determine the following commutator 
\begin{equation}
\left[ i{\bf k}_{\perp }{\bf r}_{\perp }(t),-i{\bf k}_{\perp }{\bf r}_{\perp
}(t_{1})\right] _{-}=-\frac{i\hbar k_{\perp }^{2}}{m\omega _{c}}\sin (\omega
_{c}(t-t_{1})),  \eqnum{B2}
\end{equation}
where $k_{\perp }=\left| {\bf k}_{\perp }\right| =\sqrt{k_{x}^{2}+k_{y}^{2}}$
is the magnitude of the transverse wave vector. With the operator equation
(Baker-Campbell-Hausdorff identity) 
\begin{equation}
e^{\widehat{A}}e^{\widehat{B}}=e^{\widehat{A}+\widehat{B}}e^{-\frac{1}{2}%
\left[ \widehat{A},\widehat{B}\right] _{-}},  \eqnum{B3}
\end{equation}
(which is valid, when $\left[ \widehat{A},\widehat{B}\right] _{-}$ is a
c-number) we obtain 
\begin{eqnarray}
\frac{i}{\hbar }\left[ X_{{\bf k}}(t),X_{-{\bf k}}(t_{1})\right] _{-} &=&%
\frac{2}{\hbar L^{3}}f^{2}(k_{z})\exp \left\{ i{\bf k}_{\perp }\left( {\bf r}%
_{\perp }{\bf (}t{\bf )-r}_{\perp }{\bf (}t_{1}{\bf )}\right) \right\} \sin
\left( \frac{\hbar k_{\perp }^{2}}{2m\omega _{c}}\sin (\omega
_{c}(t-t_{1}))\right) ,  \eqnum{B4} \\
\frac{1}{2}\left[ X_{{\bf k}}(t),X_{-{\bf k}}(t_{1})\right] _{+} &=&\frac{1}{%
L^{3}}f^{2}(k_{z})\exp \left\{ i{\bf k}_{\perp }\left( {\bf r}_{\perp }{\bf (%
}t{\bf )-r}_{\perp }{\bf (}t_{1}{\bf )}\right) \right\} \cos \left( \frac{%
\hbar k_{\perp }^{2}}{2m\omega _{c}}\sin (\omega _{c}(t-t_{1}))\right) . 
\nonumber
\end{eqnarray}
We assume the coordinate fluctuations to be approximately Gaussian (see Ref.
\cite{Mourokh} for the corresponding discussion) and, consequently, obtain 
\begin{equation}
\left\langle \exp \left\{ i{\bf k}_{\perp }\left( {\bf r}_{\perp }{\bf (}t%
{\bf )-r}_{\perp }{\bf (}t_{1}{\bf )}\right) \right\} \right\rangle =\exp
\left\{ -\frac{1}{2}\left\langle \left( {\bf k}_{\perp }\left( {\bf r}%
_{\perp }{\bf (}t{\bf )-r}_{\perp }{\bf (}t_{1}{\bf )}\right) \right)
^{2}\right\rangle \right\}  \eqnum{B5}
\end{equation}
and 
\begin{eqnarray}
&&\exp \left\{ i{\bf k}_{\perp }\left( {\bf r}_{\perp }{\bf (}t{\bf )-r}%
_{\perp }{\bf (}t_{1}{\bf )}\right) \right\} -\left\langle \exp \left\{ i%
{\bf k}_{\perp }\left( {\bf r}_{\perp }{\bf (}t{\bf )-r}_{\perp }{\bf (}t_{1}%
{\bf )}\right) \right\} \right\rangle  \eqnum{B6} \\
&\approx &i{\bf k}_{\perp }\left( {\bf r}_{\perp }{\bf (}t{\bf )-r}_{\perp }%
{\bf (}t_{1}{\bf )}\right) \left\langle \exp \left\{ i{\bf k}_{\perp }\left( 
{\bf r}_{\perp }{\bf (}t{\bf )-r}_{\perp }{\bf (}t_{1}{\bf )}\right)
\right\} \right\rangle .  \nonumber
\end{eqnarray}
With these simplifications we have 
\begin{equation}
G_{x,y}\left[ V_{x}(t);V_{y}(t)\right] -\left\langle G_{x,y}\left[
V_{x}(t);V_{y}(t)\right] \right\rangle =  \eqnum{B7}  \label{3}
\end{equation}
\begin{eqnarray*}
&=&-\frac{1}{mL^{3}}\sum_{{\bf k}}ik_{x,y}f^{2}(k_{z})\int_{-\infty
}^{t}dt_{1}\left\{ \left( M_{{\bf k}}(t;t_{1})+\Phi _{{\bf k}}\right) \frac{2%
}{\hbar }\sin \left( \frac{\hbar k_{\perp }^{2}}{2m\omega _{c}}\sin (\omega
_{c}(t-t_{1}))\right) +\right. \\
&&\left. +\varphi _{{\bf k}}(t;t_{1})\cos \left( \frac{\hbar k_{\perp }^{2}}{%
2m\omega _{c}}\sin (\omega _{c}(t-t_{1}))\right) \right\} i{\bf k}_{\perp
}\left( {\bf r}_{\perp }{\bf (}t{\bf )-r}_{\perp }{\bf (}t_{1}{\bf )}\right)
\left\langle \exp \left\{ i{\bf k}_{\perp }\left( {\bf r}_{\perp }{\bf (}t%
{\bf )-r}_{\perp }{\bf (}t_{1}{\bf )}\right) \right\} \right\rangle .
\end{eqnarray*}
We make the further assumption that the correlation time of the
electron-phonon interaction, $\tau _{c}$, is much less than the period of
the cyclotron oscillations, i.e. $\omega _{c}\tau _{c}<<1$ (the same
approximation was used in Ref. \cite{Ivchenko}). This is reasonable for
semiconductors at moderate magnetic fields and not too low temperatures, and
it leads to 
\begin{equation}
{\bf k}_{\perp }\left( {\bf r}_{\perp }{\bf (}t{\bf )-r}_{\perp }{\bf (}t_{1}%
{\bf )}\right) \approx k_{x}V_{x}(t)(t-t_{1})+k_{y}V_{y}(t)(t-t_{1}), 
\eqnum{B8}
\end{equation}
\begin{equation}
\frac{i\hbar k_{\perp }^{2}}{2m\omega _{c}}\sin (\omega
_{c}(t-t_{1}))\approx \frac{i\hbar k_{\perp }^{2}}{2m}(t-t_{1}),  \eqnum{B9}
\end{equation}
and 
\begin{equation}
\left\langle \exp \left\{ i{\bf k}_{\perp }\left( {\bf r}_{\perp }{\bf (}t%
{\bf )-r}_{\perp }{\bf (}t_{1}{\bf )}\right) \right\} \right\rangle =\exp
\left\{ -\frac{\left( t-t_{1}\right) ^{2}}{2\tau _{c}^{2}({\bf k}_{\perp })}%
\right\} ,  \eqnum{B10}
\end{equation}
where 
\begin{equation}
\tau _{c}^{-2}({\bf k}_{\perp })=k_{x}^{2}\left\langle
V_{x}^{2}(t)\right\rangle +k_{y}^{2}\left\langle V_{y}^{2}(t)\right\rangle
+k_{x}k_{y}\left\langle \left[ V_{x}(t),V_{y}(t)\right] _{+}\right\rangle . 
\eqnum{B11}
\end{equation}
The resulting simplified stochastic equations for the fluctuating velocity
components take the form given by Eq.(\ref{Velocity}).

\begin{figure}[tbp]
\caption{Dependencies of the relaxation rate on the energy of the lowest
level of the quantum well; (a) for temperature $T=40K$ and two magnetic
field strengths, (b) for temperature $T=300K$ (the same curve for both
magnetic field strengths).}
\label{fig1}
\end{figure}
\begin{figure}[tbp]
\caption{ Dependence of the relaxation rate on magnetic field strength for
various temperatures.}
\label{fig2}
\end{figure}
\begin{figure}[tbp]
\caption{Temperature dependence of the relaxation rate for various magnetic
field strengths.}
\label{fig3}
\end{figure}

\end{document}